\documentstyle[preprint,aps]{revtex}

\begin{document}
\draft
\preprint{KANAZAWA 96-11, August 1996}
\title{
Monopole action and monopole condensation \\
in SU(3) lattice QCD 
}
\author{
Natsuko Arasaki$^{\small a}$, Shinji Ejiri$^{\small a}$, 
Shun-ichi Kitahara$^{\small b}$,\\
 Yoshimi Matsubara$^{\small c}$ 
 and Tsuneo Suzuki$^{\small a}$
\footnote{
E mail address: suzuki@hep.s.kanazawa-u.ac.jp}
}
\address{$^{\small a}$
Department of Physics, Kanazawa University, Kanazawa 920-11, Japan
}
\address{$^{\small b}$
Jumonji University, Niiza , Saitama 352,  Japan
}
\address{$^{\small c}$
Nanao Junior College, Nanao, Ishikawa 926, Japan
}
\date{\today}
\maketitle
\begin{abstract}
Effective monopole actions for various extended monopoles 
are derived from vacuum configurations 
after abelian projection in the maximally abelian gauge in 
$T=0$ and $T\ne 0$ $SU(3)$ lattice QCD. 
The actions obtained appear to be independent 
of the lattice volume adopted. 
At zero temperature, 
monopole condensation is seen to occur from energy-entropy balance 
in the strong coupling region.
 Larger $\beta$ is included in the monopole condensed phase as more 
extended monopoles are considered. 
The scaling seen in the $SU(2)$ case is not yet observed. 
The renormalization flow diagram suggests the existence of an infrared 
fixed point.
A hysteresis behavior is seen around the critical 
temperature in the case of the $T\ne 0$ action.
\end{abstract}


\newpage
\narrowtext

\section{Introduction}
The dual Meissner effect is believed to be the
promising candidate for the quark confinement 
mechanism~\cite{thooft1,mandel}. 
This picture is realized in the confinement phase of lattice compact
QED\cite{poly,bank,degrand}.  
In QCD, 
the 'tHooft idea \cite{thooft2} 
of abelian projection is very interesting.
The abelian projection is to extract
an abelian gauge theory by performing a partial gauge-fixing.
Abelian projected QCD can be regarded as an abelian theory with electric 
charges and magnetic monopoles. 'tHooft conjectured that the condensation 
of the abelian monopoles causes the confinement in QCD.

Many works have been done to test the idea in the framework of lattice QCD
using Monte-Carlo simulations. An interesting abelian projection called 
maximally abelian (MA) gauge \cite{kron} is found.
$U(1)^2$ invariant operators written in terms of abelian link 
fields alone after the abelian 
projection reproduce essential features of confinement phenomena like 
the string tension \cite{yotsu}, the Polyakov loop, thermodynamic 
quantities \cite{hio91,suzu93} and even chiral condensate and 
hadron masses \cite{wolo94,miya95,wolo95,suzu95a}. 
This is called abelian dominance.

Such $U(1)^2$ invariant abelian operators can be decomposed into 
a product of a monopole operator written 
in terms of monopole currents or Dirac string and a photon one containing 
photon contribution alone \cite{shiba94,stack94,suzu95b}.
The above phenomena called abelian dominance are shown to be reproduced
by the monopole contributions alone (monopole dominance) 
\cite{miya95,wolo95,suzu95a,shiba94,stack94,suzu95b,ejiri95a}.
 
Such phenomena called abelian and monopole dominance 
strongly suggest that low-energy QCD can be described by an effective 
abelian theory. Actually it is possible to derive an effective theory 
of an extended Wilson form
in terms of the abelian link field alone \cite{suzu93}. But the action derived 
needs larger and larger Wilson loops when we go to higher $\beta$ in the 
scaling region, although it takes a simple Wilson action like in compact QED 
in the strong coupling 
region.

The monopole dominance implies the existence of an effective action on the 
dual lattice in terms of a dual quantity like monopole currents. 
In the case of compact QED, the exact dual transformation can be done and 
leads us to an action describing a monopole Coulomb gas,
when one adopt the partition function of the Villain form
\cite{bank,peskin,frolich,smit}. 
Monopole condensation is shown to 
occur in the confinement phase from energy-entropy balance
of monopole loops. 

In the case of QCD, however, we encounter a difficulty 
in performing the exact dual transformation. 
Shiba and 
one of the present author (T.S.) have succeeded in 
numerically carrying out the dual transformation to obtain 
a monopole action from the vacuum ensemble  
of monopole currents \cite{shiba1,shiba2,shiba3,shiba4,shiba5}. 
This can be done by extending the Swendsen method\cite{swendsen}.
They also have performed a block-spin transformation on the dual lattice by 
considering extended monopoles \cite{ivanenko}.
The monopole action determined in $SU(2)$ show the following interesting 
behaviors:
\begin{enumerate}
\item A compact and local form of the monopole action is obtained 
even in the scaling region. The coupling constant of the self-interaction is 
dominant and the coupling constants decrease rapidly as the distance 
between the two monopole currents increases.
\item Coupling constants $f_i$ for any effective action 
look volume independent.
\item Monopole condensation is seen to occur for smaller $\beta$ from 
energy-entropy balance.
\item $f_i$ look to show a scaling behavior, that is, they  
are written only by a physical scale defined by $b=n a(\beta)$.
This suggests that the $SU(2)$ monopole action is near 
the renormalized trajectory 
of the block spin transformation.
\item If the scaling holds good 
even on the 
infinite lattice, the $SU(2)$ QCD vacuum is always (for all $\beta$) 
in the monopole condensed and then  confined phase.
\end{enumerate}

To extend this method to $SU(3)$ QCD is very interesting, but it is 
not so straightforward.  There are two independent 
monopole currents and, to speak more rigorously, three currents 
$k^i_{\mu}(s)$ satisfying 
one constraint $\sum_i k^i_{\mu}(s)=0$. There is a permutation symmetry 
(Weyl symmetry) with respect to the species.  Calculating the entropy becomes 
very difficult as naturally expected. Hence we try to construct an 
effective action composed of only one monopole current after integrating 
out the other two. Then the entropy may be evaluated similarly as done in 
$SU(2)$ and in compact QED.
It is the aim of this note to report the results 
of $SU(3)$ QCD\cite{shiba5,suzu96}. 

\section{MA gauge and monopole currents in $SU(3)$}
The MA gauge is given on a lattice  by performing
a local gauge transformation 
	\begin{equation}
		\widetilde{U}_{\mu}(s)=V(s)U_{\mu}(s)V^{-1}(s+\hat\mu)
	\end{equation}
such that a quantity
	\begin{eqnarray}
		R &=& \sum_{s,\mu,a}{\rm Tr}
		\Big(U_{\mu}(s)\,\lambda_a\,
		U^{\dagger}_{\mu}(s)\,\lambda_a\Big),
	\end{eqnarray}
	\vspace{-5mm}
	\begin{eqnarray}
        \lambda_1 &=& \left(
			\begin{array}{ccc}
			1 & 0 & 0 \\
			0 & -1 & 0 \\
			0 & 0 & 0
			\end{array}
		\right),
        \lambda_2= \left(
			\begin{array}{ccc}
			-1 & 0 & 0 \\
			0 & 0 & 0 \\
			0 & 0 & 1
			\end{array}
		\right),
        \lambda_3= \left(
			\begin{array}{ccc}
			0 & 0 & 0 \\
			0 & 1 & 0 \\
			0 & 0 & -1
			\end{array}
		\right)\nonumber
	\end{eqnarray}
is maximized. Then
a quantity 
	\begin{eqnarray}
		X(s) = \sum_{\mu,a}\,[(U_{\mu}(s)\,\lambda_a\,
			U^{\dagger}_{\mu}(s)+U^{\dagger}_{\mu}(s-\mu)\,
			\lambda_a\,U_{\mu}(s-\mu)),\lambda_a] \label{X}
	\end{eqnarray}
vanishes.
After the gauge fixing is done, an abelian link gauge field $u_{\mu}(s)$ 
is extracted from $SU(3)$ link variables as follows\cite{schier};
	\begin{eqnarray}
		\tilde U_{\mu}(s) &\equiv& C_{\mu}(s)\,u_{\mu}(s),\\
		\nonumber\\
		u_{\mu}(s) &=& \left(
			\begin{array}{ccc}
			e^{i\theta_{\mu}^1(s)} & 0 & 0 \\
			0 & e^{i\theta_{\mu}^2(s)} & 0 \\
			0 & 0 & e^{i\theta_{\mu}^3(s)} 
			\end{array}
		\right),\ \ \ 
		\sum_{i=1}^{3}\,\theta_{\mu}^i(s) = 0\hspace{.2cm} 
                 ({\textstyle\rm mod 2\pi}),
	\end{eqnarray}
where
	\begin{eqnarray}
		\theta_{\mu}^i(s) &\equiv& arg\,([\tilde U_{\mu}(s)]_{ii})
			-\frac {1}{3}\,\phi_{\mu}(s)
			\hspace*{3mm}\in (-\frac {4}{3}\pi,\frac {4}{3}\pi),\\
		\phi_{\mu}(s) &\equiv& \left.\sum_{i=1}^{3}\,
			arg\,([ \tilde U_{\mu}(s)]_{ii})\,\right |
			_{\textstyle\rm mod 2\pi}
			\in [-\pi,\pi)
	\end{eqnarray}
and the plaquette angles $\Theta_{\mu\nu}^i(s)$ are given by the sum of 
link angles $\theta_{\mu}^i$ as follows;
	\begin{eqnarray}
		\Theta_{\mu\nu}^i(s)
			&=& \partial_{\mu} \theta_{\nu}^i
			-\partial_{\nu} \theta_{\mu}^i,\nonumber\\
		\sum_{i=1}^{3}\,\Theta_{\mu\nu}^i&=&2 \pi \ell
		\hspace{3mm}(\,\ell\hspace{1mm}=\hspace{1mm}0,\pm 1).
	\end{eqnarray}
If $\ell=+1$, the plaquette phases are chosen so that
	\begin{eqnarray}
		\tilde \Theta_{\mu\nu}^i(s)=\left\{ \begin{array}{ll}
		\Theta_{\mu\nu}^i(s)-2 \pi
			&  {\rm if} \hspace{3mm} 
                        \Theta_{\mu\nu}^i(s)=\rm { max\,(\Theta_{\mu\nu}^1,
                        \Theta_{\mu\nu}^2,\Theta_{\mu\nu}^3)}\\
		\Theta_{\mu\nu}^i(s)
			& {\rm otherwise} \end{array} \right.
	\end{eqnarray} 
If $\ell=-1$,
	\begin{eqnarray}
		\tilde \Theta_{\mu\nu}^i(s)=\left\{ \begin{array}{ll}
		\Theta_{\mu\nu}^i(s)+2 \pi
			& {\rm if} \hspace{3mm} 
			\Theta_{\mu\nu}^i(s)=\rm { min\,(\Theta_{\mu\nu}^1,
                        \Theta_{\mu\nu}^2,\Theta_{\mu\nu}^3)}\\
		\Theta_{\mu\nu}^i(s)
			& {\rm otherwise}. \end{array} \right.
	\end{eqnarray}
Using  $\tilde \Theta_{\mu\nu}^i(s)$, the monopole current is given by
	\begin{eqnarray}
                 &&\bar\Theta_{\mu\nu}^i(s)
               =\tilde \Theta_{\mu\nu}^i(s)-2\pi n_{\mu\nu}^i(s)
               \hspace*{3mm}\in [-\pi,\pi),\\
	         && k_{\mu}^{i}(s) = 
		 -\frac{1}{4 \pi} \epsilon_{\mu \nu \rho \sigma}
	         \partial_{\nu} \bar \Theta_{\rho \sigma}^{i}(s+\hat\mu),\\
		 && \sum_{i=1}^{3}\,k_{\mu}^i(s) = 0.
	\end{eqnarray}
A block-spin transformation is done by considering 
$n^3$ extended monopoles \cite{ivanenko}:
	\begin{eqnarray}
 	k_{\mu}^{(n)i}(s) &=& -\frac {1}{4 \pi}\,\epsilon_{\mu\nu\rho\sigma}\,
		\partial_{\nu}\bar\Theta_{\rho\sigma}^{(n)i}(s+\hat\mu)
		\nonumber\\
		&=& \sum_{i,j,m=0}^{n-1}k_{\mu}^i (n s
		+(n-1)\hat\mu+i\hat\nu+j\hat\rho
		+m\hat\sigma)\\
	\bar\Theta_{\rho\sigma}^{(n)i}(s) &\equiv& 
		\sum_{i,j=0}^{n-1}\bar\Theta_{\rho\sigma}^i
		(n s+i\hat\rho+j\hat\sigma)
	\end{eqnarray}

\section{The method}
The dual transformation is done as follows:
\begin{eqnarray}
Z    & = &\int e^{-S(U)}\delta(X)\Delta_F(U)DU\\
    & = &\int Du [\int Dc e^{-S(u,c)}
                     \delta(X)\Delta_F(U)]\\
    & = &\int Due^{-S_{eff}(u)}\\
    &=& (\prod\sum)
 \int Du \delta(k,u) e^{-S_{eff}(u)},\\
&= &(\prod\sum)e^{-S[k^i]},
\end{eqnarray}
where 
$X$ is the quantity (\ref{X}), 
 $\Delta_F(U)$ is the Fadeev-Popov determinant and 
\begin{eqnarray}
(\prod\sum)&\equiv&(\prod_{s,\mu,i}\sum_{k^i_{\mu}(s)=-\infty}^{\infty})
(\delta_{\partial'_{\mu}k^i_{\mu}(s),0})\delta(\sum_j k^j_{\mu}(s)),
 \label{prod} \\
\delta(k,u) &\equiv& \delta(
k^i_{\mu}(s) + \frac{1}{4\pi}\epsilon_{\mu\nu\rho\sigma}
\partial_{\nu}\bar \Theta_{\rho \sigma}^{i}(s+\hat{\mu})).
\end{eqnarray}

The block-spin transformation 
on the dual lattice \cite{ivanenko} is expressed as 
\begin{eqnarray}
Z&=&(\prod\sum)'[(\prod\sum)\delta(k_{\mu}^{(n)i}(s)
 - \sum_{i,j,l=0}^{n-1}k^i_{\mu}(ns+(n-1)\hat{\mu}
+i\hat{\nu}+j\hat{\rho}+l\hat{\sigma}))
e^{-S[k^i]}]\\
&=&(\prod\sum)'e^{-S[k^{(n)i}]},
\end{eqnarray}
where $(\prod\sum)'$ is defined for the extended currents similary
as in (\ref{prod}).

\begin{figure}[htb]
            \vspace*{3cm}
		\epsfxsize=\textwidth
		\centerline{\epsffile{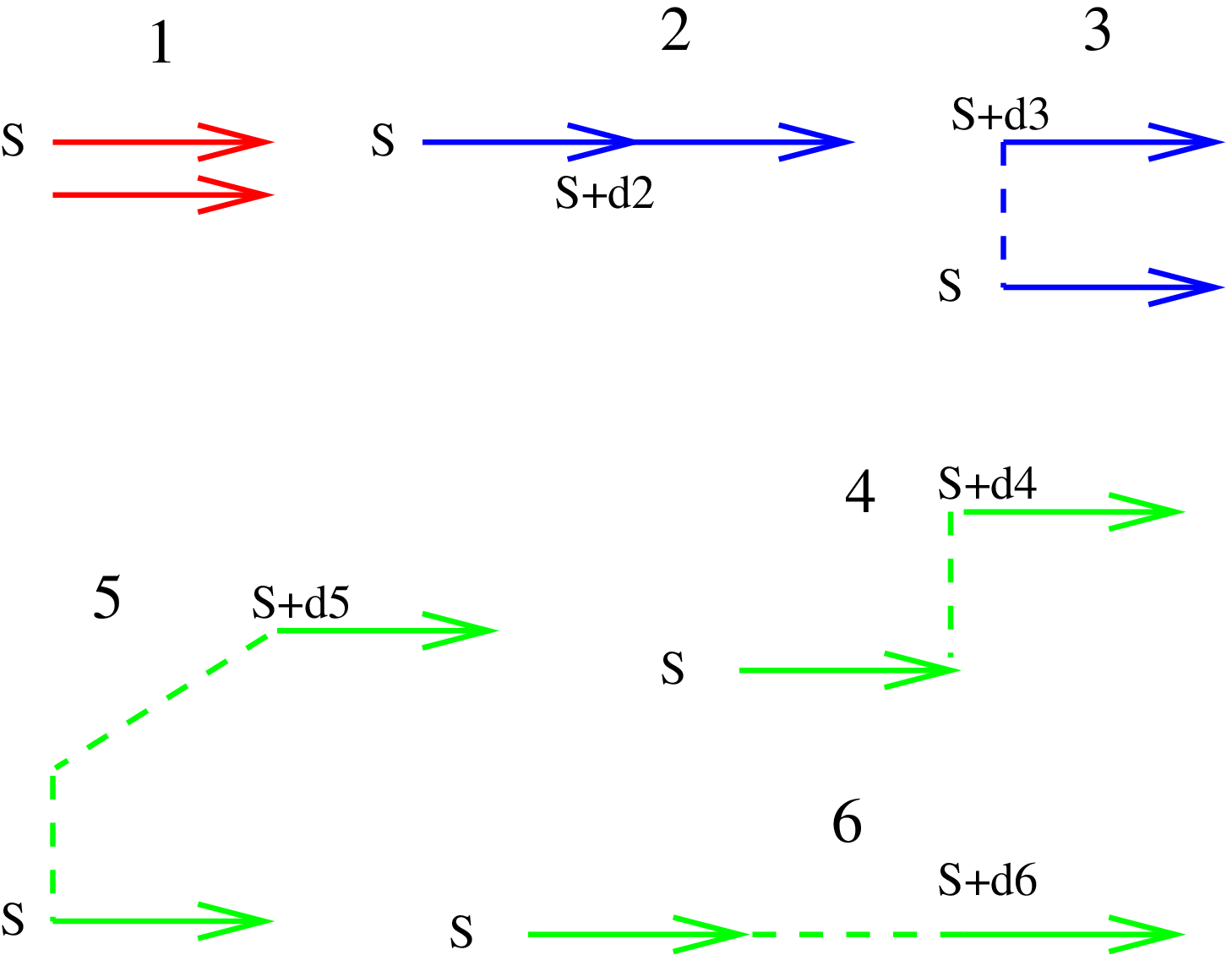}}
            \vspace*{4cm}
		\caption{
		The first six terms of monopole interactions in the action.
}
		\label{current}
	\end{figure}

As shown above, we try to fix the monopole action after integrating out 
two monopole currents.
The monopole action adopted is composed of various two current
interactions $S[k]=\sum_i\,f_i\,S_i[k]$. 
Practically we have to restrict the number of 
interaction terms. We  adopted
12 types of quadratic interactions
in most of these studies as was done in $SU(2)$.
The definitions are 
listed in \cite{shiba1,shiba4}. Here we show the first important 6 
interactions in Fig.\ \ref{current}.

%
%
In the $T \ne 0$ QCD, the 
time extent is finite. Hence, the monopole action is taken as follows:
	\begin{eqnarray}
		S[k]= \sum_i\,(f_i^s\,S_ {i,s}+f_i^t\,S_{i,t}),
		\nonumber
	\end{eqnarray}
where $S_{i,s}$ are interactions between space-like currents and
$S_{i,t}$ are interactions between time-like currents.

We generate thermalized vacuum configurations $U_{\mu}(s)$
and then perform the partial gauge fixing in the MA gauge. Then 
using the above definition of the monopole and 
the extended monopoles, we get the vacuum ensemble of 
$k_{\mu}(s)$ and $k_{\mu}^{(n)}(s)$ 
currents. The Swendsen method\cite{swendsen} 
 is applied to these current ensembles.
Since the dynamical variables $k_{\mu}(s)$ satisfy the 
conservation rule, it is necessary to extend the 
original Swendsen method by 
considering a plaquette $(s',\mu',\nu')$ instead of a link
\cite{shiba4,shiba6}.
Introducing a new set of coupling constants 
$ \{\tilde{f}_i\} $, define
\begin{equation}
\bar{S}_i [k]= \frac{\sum_{M=-\infty}^{\infty}S_i[k']
\exp(-\sum_j \tilde{f}_j S_j [k'])}
{\exp(-\sum_j \tilde{f}_j S_j [k'])},
\label{sbar}
\end{equation}
where 
$k'_{\mu}(s)  =  k_{\mu}(s) + M(\delta_{s,s'}\delta_{\mu,\mu'}
+ \delta_{s,s'+\hat{\mu}'}\delta_{\mu,\nu'}  
- \delta_{s,s'+\hat{\nu}'}\delta_{\mu,\mu'}
- \delta_{s,s'}\delta_{\mu,\nu'}).$
When all $\tilde{f}_j$ are equal to $ f_j $,
one can prove an equality   
$ \langle \bar{S}_i \rangle  =  \langle S_i \rangle $,
where the expectation values are taken 
over the above original action 
with the coupling constants $\{f_i\}$. 
Otherwise, one may expand the difference as follows: 
\begin{equation}
\langle \bar{S}_i - S_i \rangle = 
\sum_j 
\langle \overline{S_i S_j}-\bar{S}_i\bar{S}_j \rangle
(f_j - \tilde{f}_j)    \label{ssbar}, 
\end{equation}
where only the first-order terms are written down.
This allows an iteration scheme for determination 
of the unknown constants
$f_i$. For details, see the references\cite{shiba4,shiba6}.

\section{The results in the $T=0$ case.}

Since we are restricted to the one-current case, the same method can be 
applied as in $SU(2)$ QCD\cite{shiba1,shiba4}. 
The lattice sizes and $\beta$ 
considered are from $8^4$ to 
$24^4$ and from $\beta=5.0$ to $\beta=6.5$ for the $T=0$ case. 
Extended monopoles from $1^3$ to $4^3$ are studied for $T=0$.
After the thermalization, 50 configurations in the case of $24^4$ lattice 
are used for the average.
The monopole action in $SU(3)$ QCD is obtained beautifully.

\begin{figure}[htb]
		\epsfxsize=0.9\textwidth
		\centerline{\epsffile{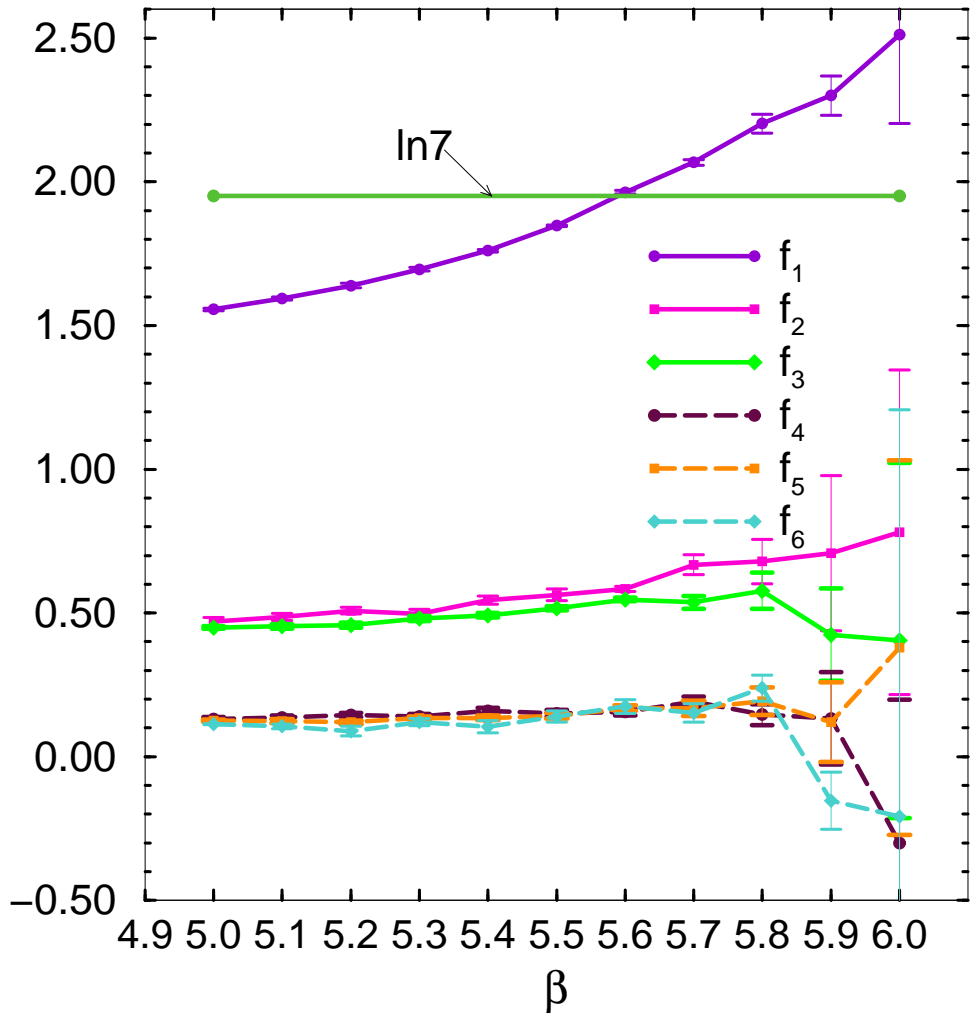}}
		\caption{
		The effective monopole action in $T=0$ $SU(3)$ QCD.
		}
		\label{fb6}
	\end{figure}

\begin{figure}[htb]
		\epsfxsize=0.9\textwidth
		\centerline{\epsffile{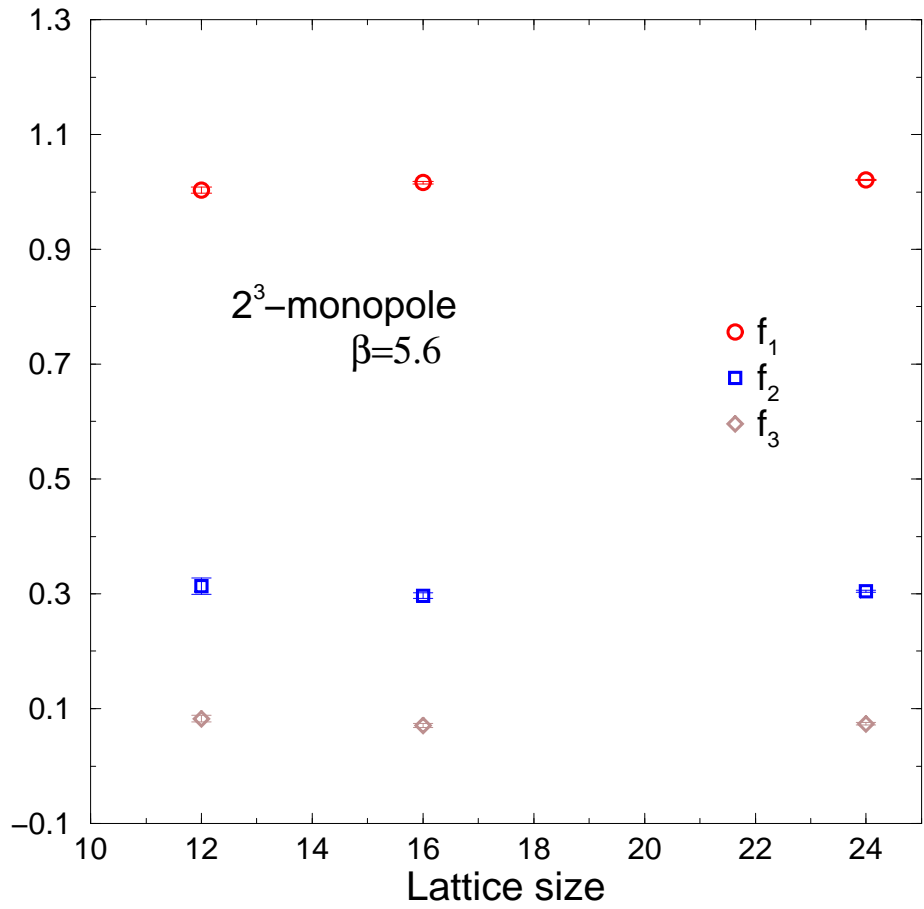}}
		\caption{
		Coupling constants $f_i$ versus lattice size.
		}
		\label{lat}
	\end{figure}

\begin{figure}[htb]
		\epsfxsize=0.9\textwidth
		\centerline{\epsffile{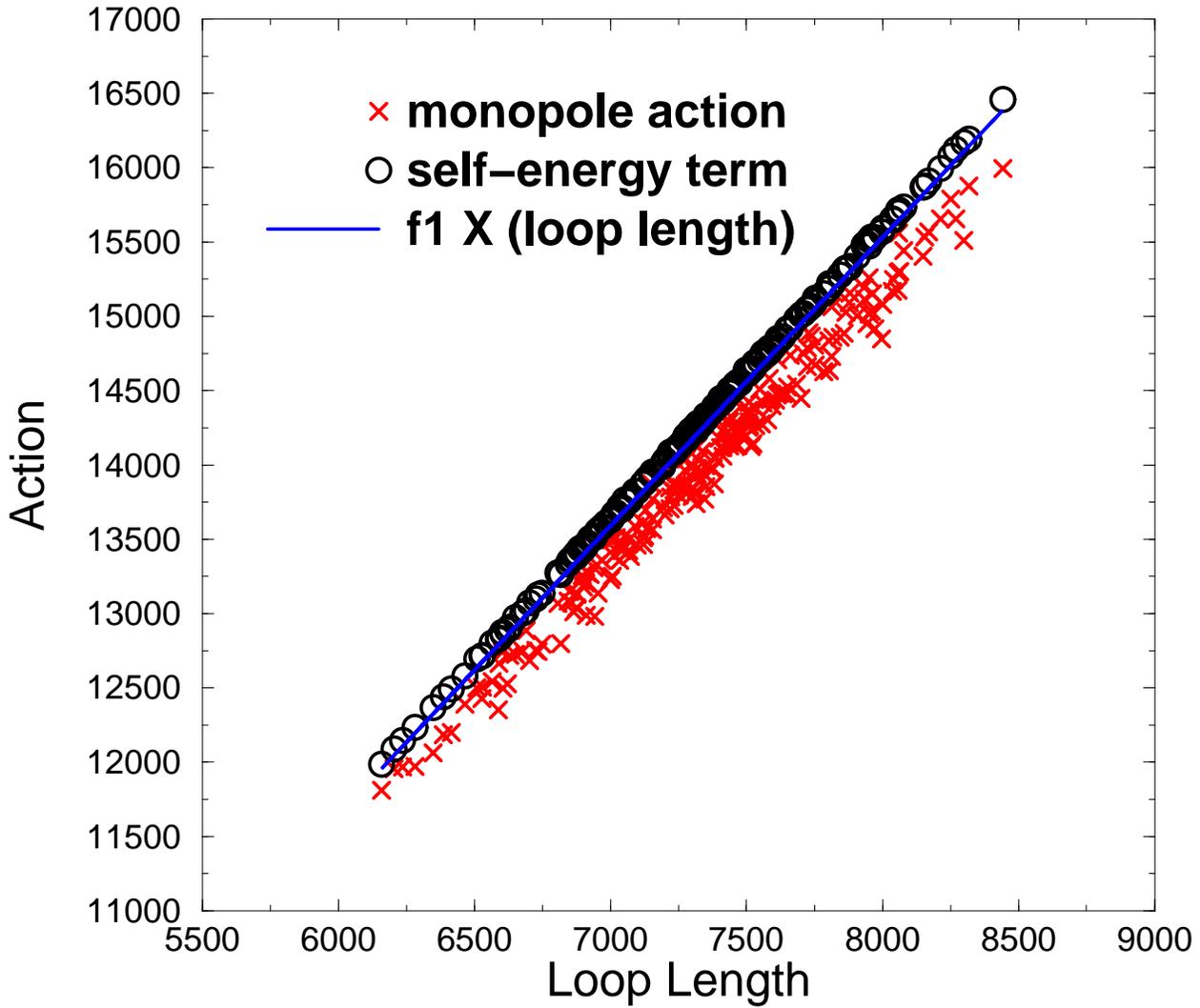}}
		\caption{
		The total monopole action, the self-energy term of 
		the action and $f_1$ times the length of a monopole loop $L$ 
		versus length of the long monopole loop.
		}
		\label{f1l}
	\end{figure}

\begin{figure}[htb]
		\epsfxsize=0.9\textwidth
		\centerline{\epsffile{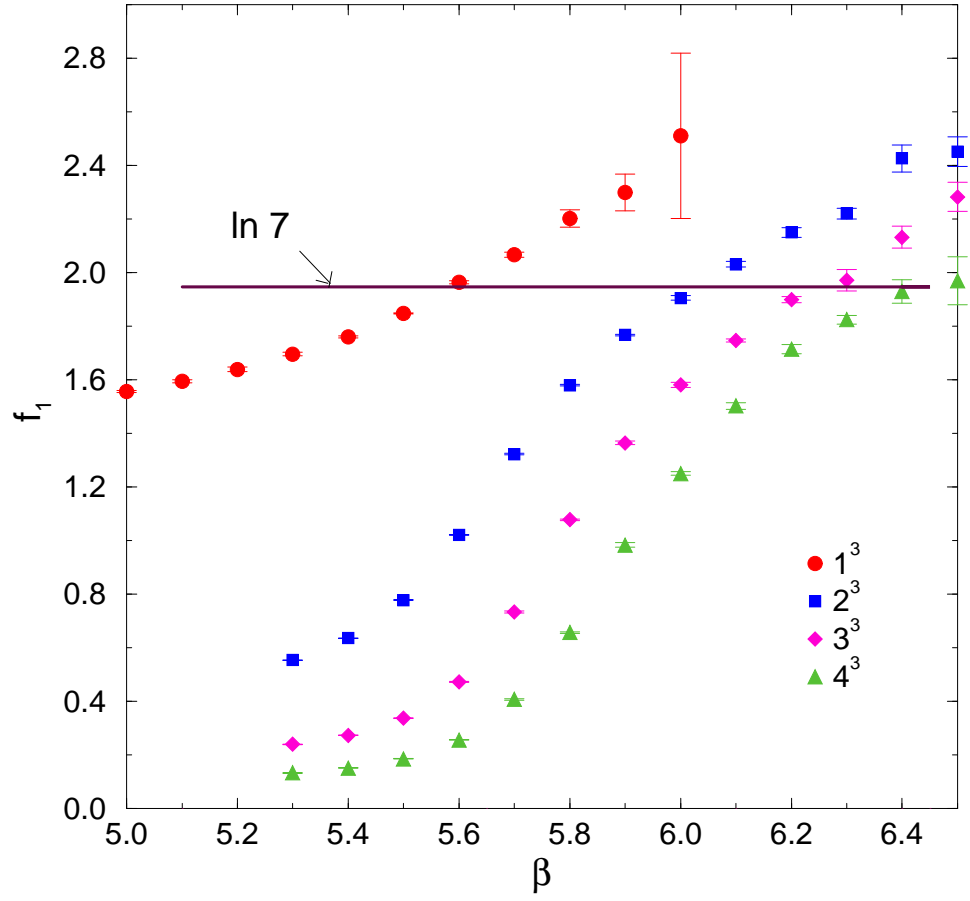}}
		\caption{
		Coupling constants $f_1$ versus $\beta$ for $2^3,3^3$ and 
		$4^3$ extended monopoles on $24^3$ lattice.
		}
		\label{f1be}
	\end{figure}

\begin{figure}[htb]
		\epsfxsize=0.9\textwidth
		\centerline{\epsffile{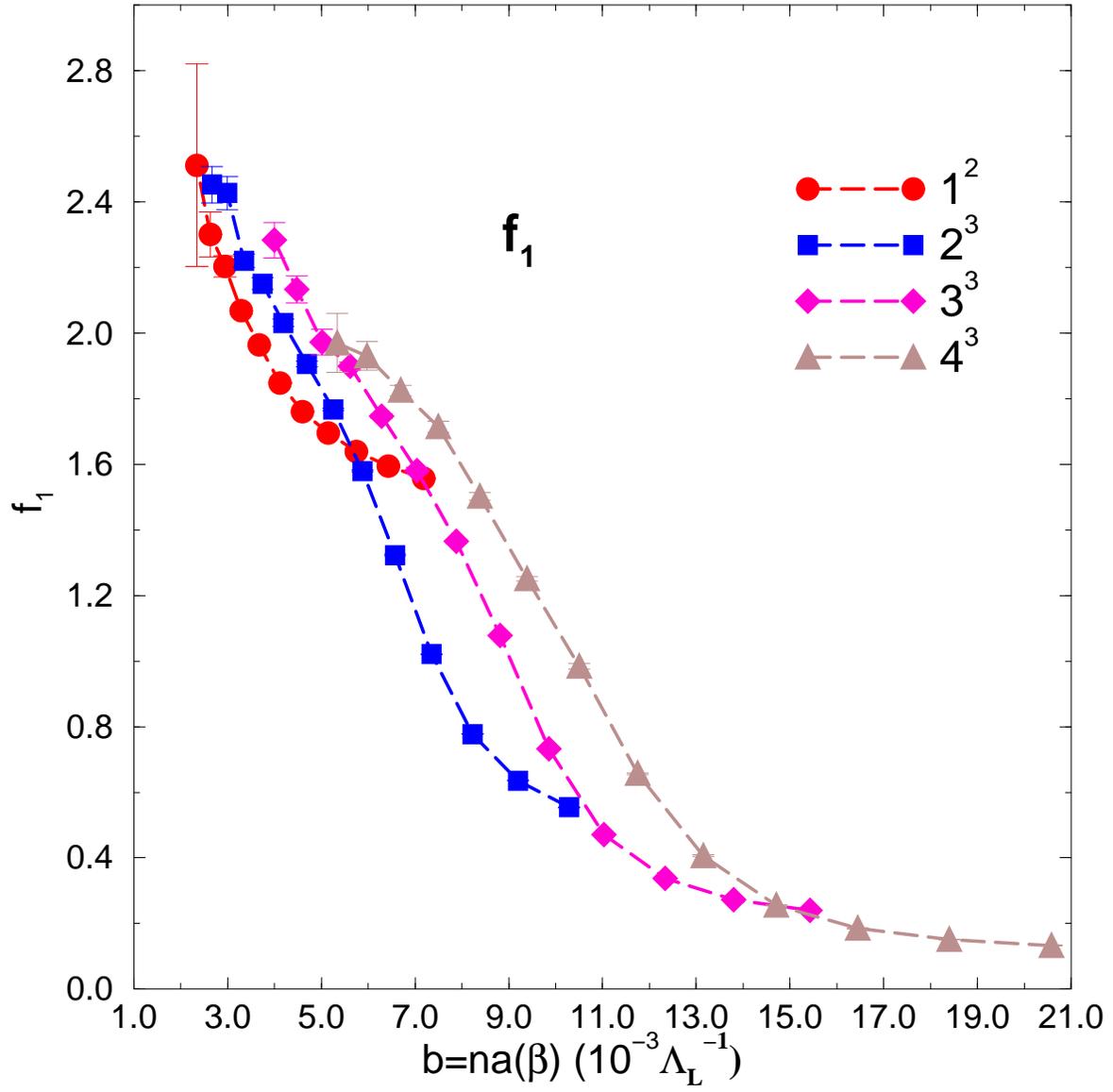}}
		\caption{
		Coupling constants $f_1$ versus $b$.
		}
		\label{f1b}
	\end{figure}

\begin{figure}[htb]
		\epsfxsize=0.9\textwidth
		\centerline{\epsffile{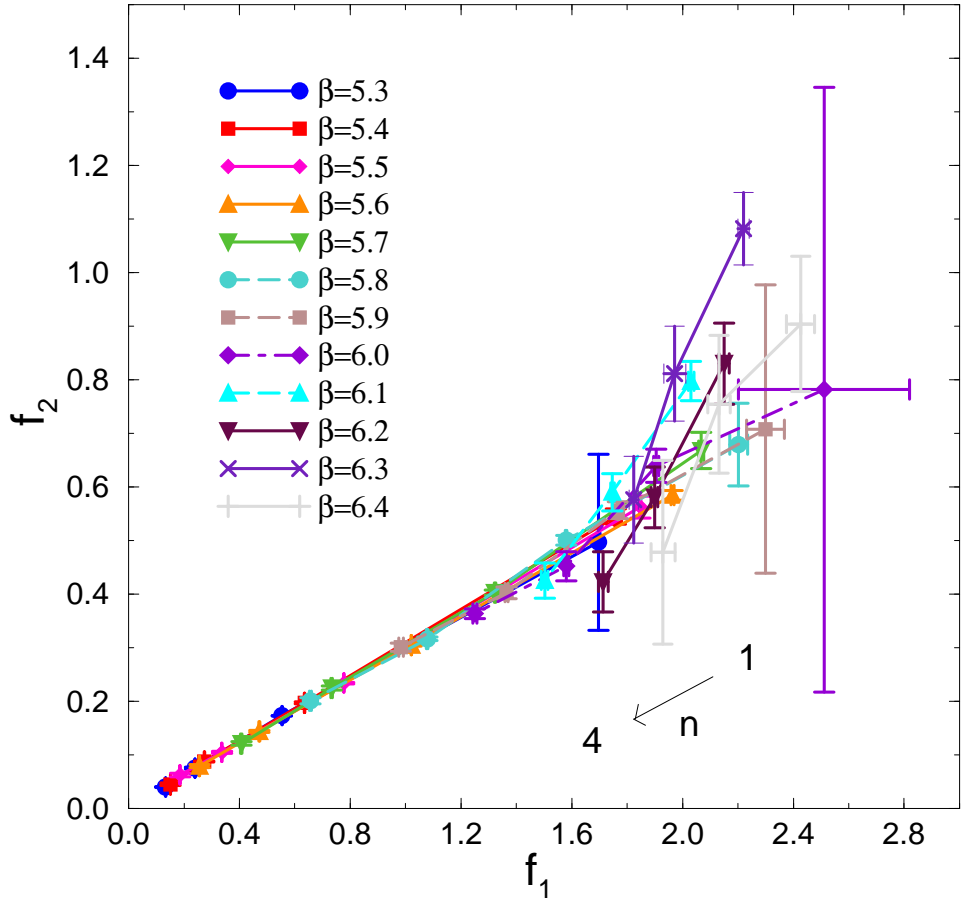}}
		\caption{
		The $f_1-f_2$ plane of the renormalization flow.
}
		\label{f12}
	\end{figure}

\begin{enumerate}
\item
The monopole actions for all extended monopoles are
 fixed in a compact form even in the 
scaling region. The self-energy term is dominant 
and the coupling constants decrease rapidly as the distance between 
the two monopole currents increase as seen in Fig.\ \ref{fb6}. 
	\begin{eqnarray}
		f_1 \gg f_2 \sim f_3 > f_4 \sim f_5 \sim f_6
	\nonumber
	\end{eqnarray}
\item
Fig.\ \ref{lat} shows the volume dependence of the typical 
action obtained.
To be stressed is that there is almost no lattice-volume dependence.
This is very interesting, since it suggests finite lattice-size  effects 
are very small.
\item
Monopole loops exist as a closed loop in the four-dimensional space.
It is found that there is a long connected loop and some short loops 
in the confinement phase, whereas only the short loops exist
in the deconfinement phase. It is known in $SU(2)$ that only the long loop 
is responsible for confinement\cite{ejiri95a}.
Hence we plot the value of the action and that of the self-energy term alone
versus the length of the long monopole loops in Fig.\ \ref{f1l}.
Although the figure is in the $T\ne 0$ case, 
the same behaviors are seen also in the $T=0$ case.
The total action is well approximated by the product of the self-coupling 
constant and the length $f_1\times L$.
\item
Since the action is well approximated by $f_1\times L$,
we plot $f_1$ versus $\beta$ for various extended monopole on 
$24^4$ lattice in Fig.\ \ref{f1be}. 
Assuming that the entropy is estimated as in compact QED,
we also show 
the entropy value $\ln 7$ per unit monopole length 
in comparison.
Each extended monopole has its own $\beta$ region 
where the condition $f_1 < \ln 7$ is satisfied. 
Since the entropy dominates over the energy, the monopole condensation occurs
also in $SU(3)$ QCD for such a $\beta$ region.
When the extendedness 
is bigger, larger  $\beta$ is included in such a region.
\item
The results obtained above are very similar to those in $SU(2)$ case.
In $SU(2)$, there is a very interesting scaling behavior. That is, 
the coupling constants are described by a physical length 
$b=na(\beta)$ where $n$ is the number of blocking and the two-loop 
perturbation value is used for $a(\beta)$.
Unfortunately, such a scaling is not yet seen in $SU(3)$ as 
shown in Fig.\ \ref{f1b}. We need to perform more steps of 
the block-spin transformations.
\item
In Fig.\ \ref{f12}, we plot the  $f_1-f_2$ plane of the renormalization 
flow. The flow line for smaller $\beta$ regions is beautifully straight with
very small errors. The slope is fixed to be $f_2/f_1\sim 0.3$.
Moreover there seems to be an infrared fixed-point at $f_i=0$.
\end{enumerate}


\section{The results in the $T\ne 0$ case.}
\begin{figure}[htb]
		\epsfxsize=0.9\textwidth
		\centerline{\epsffile{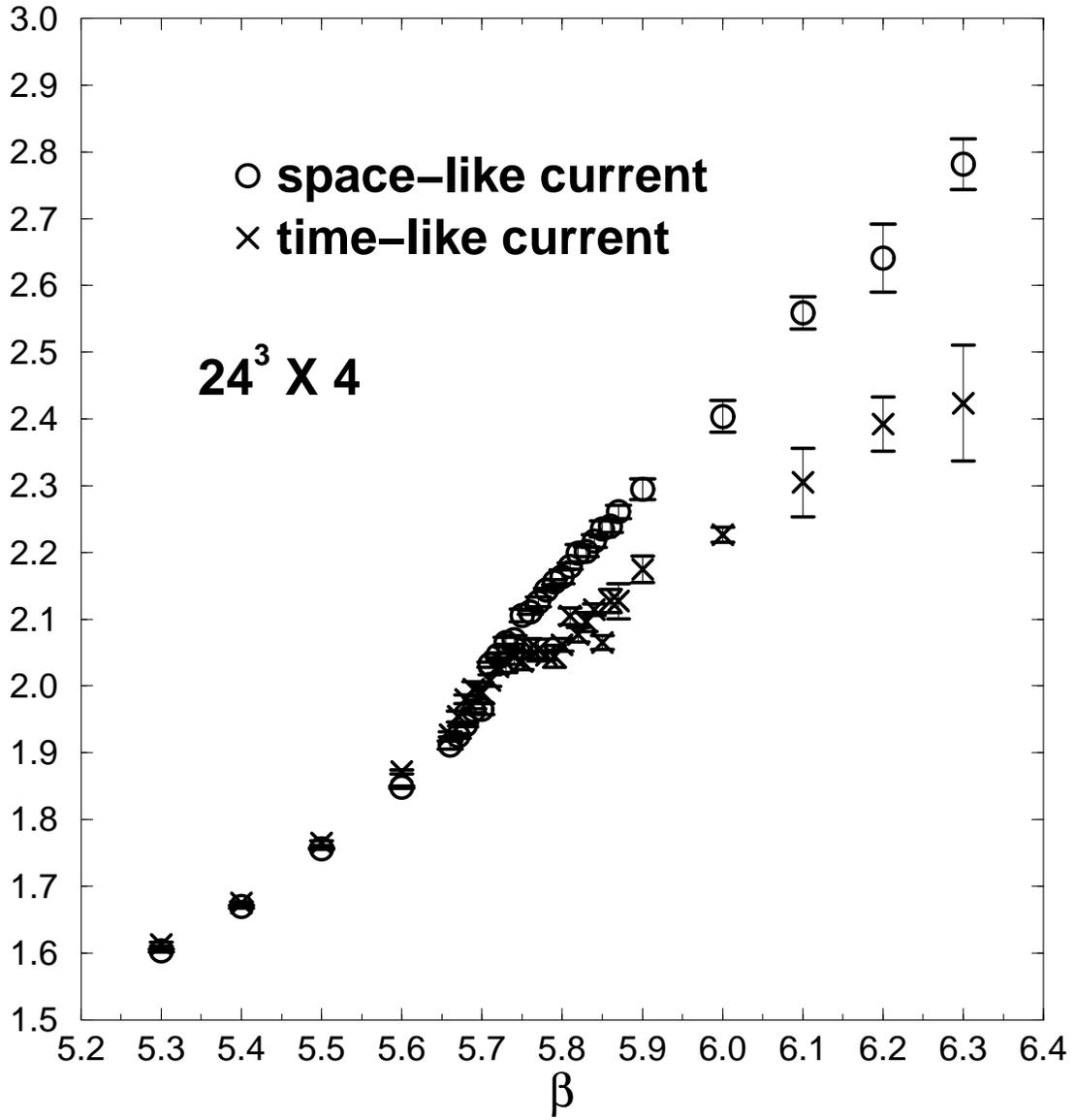}}
		\caption{
		The self-energy coupling constants for
		space-like monopole currents and for 
		time-like monopole currents in the case of hot-start. 
		}
		\label{f1st}
	\end{figure}

\begin{figure}[htb]
		\epsfxsize=0.9\textwidth
		\centerline{\epsffile{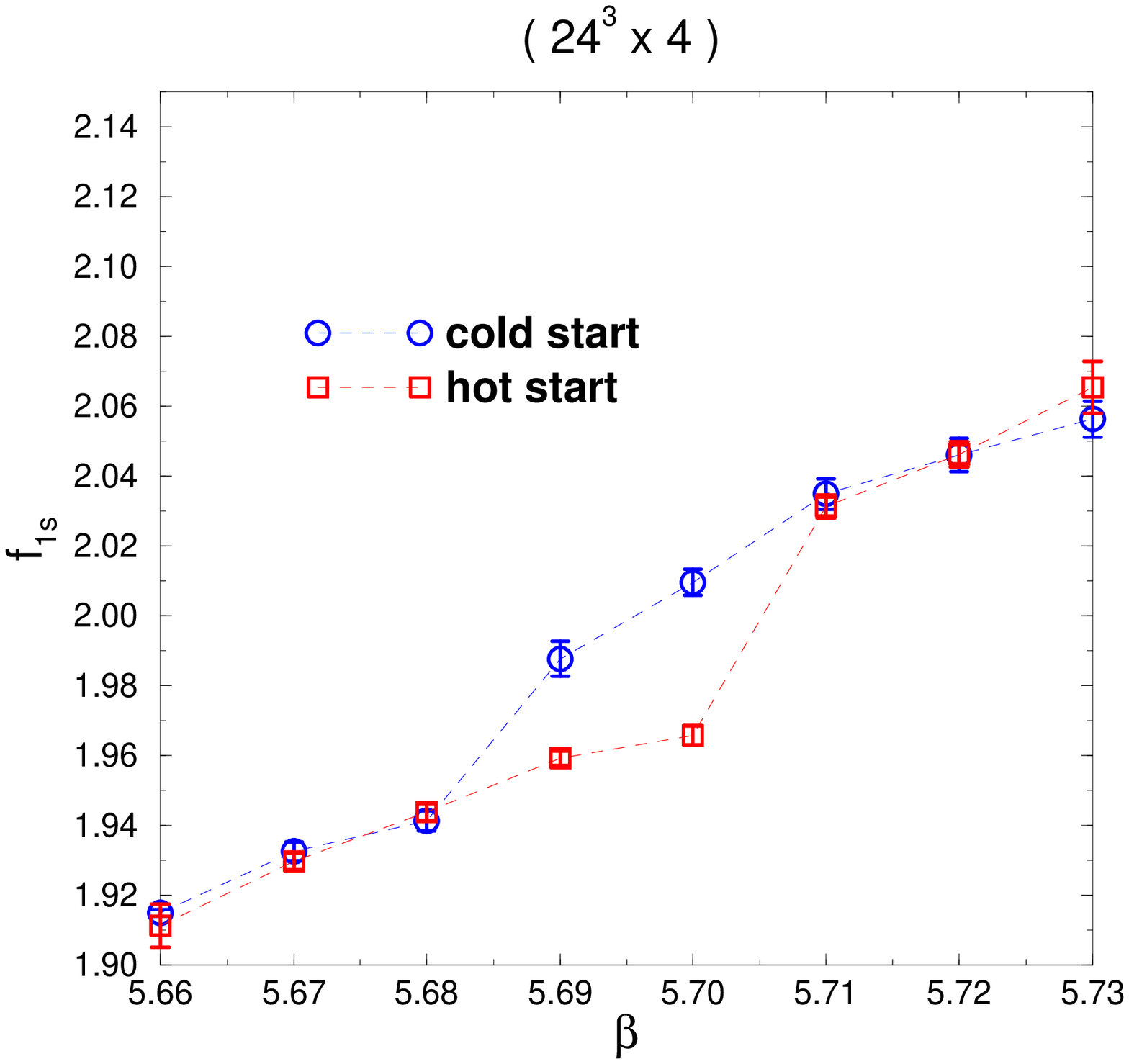}}
		\caption{
		$f_1$ for the space-like currents 
 near the transition temperature ($\beta_c = 5.69$).
		}
		\label{f1s}
	\end{figure}

\begin{figure}[htb]
		\epsfxsize=0.9\textwidth
		\centerline{\epsffile{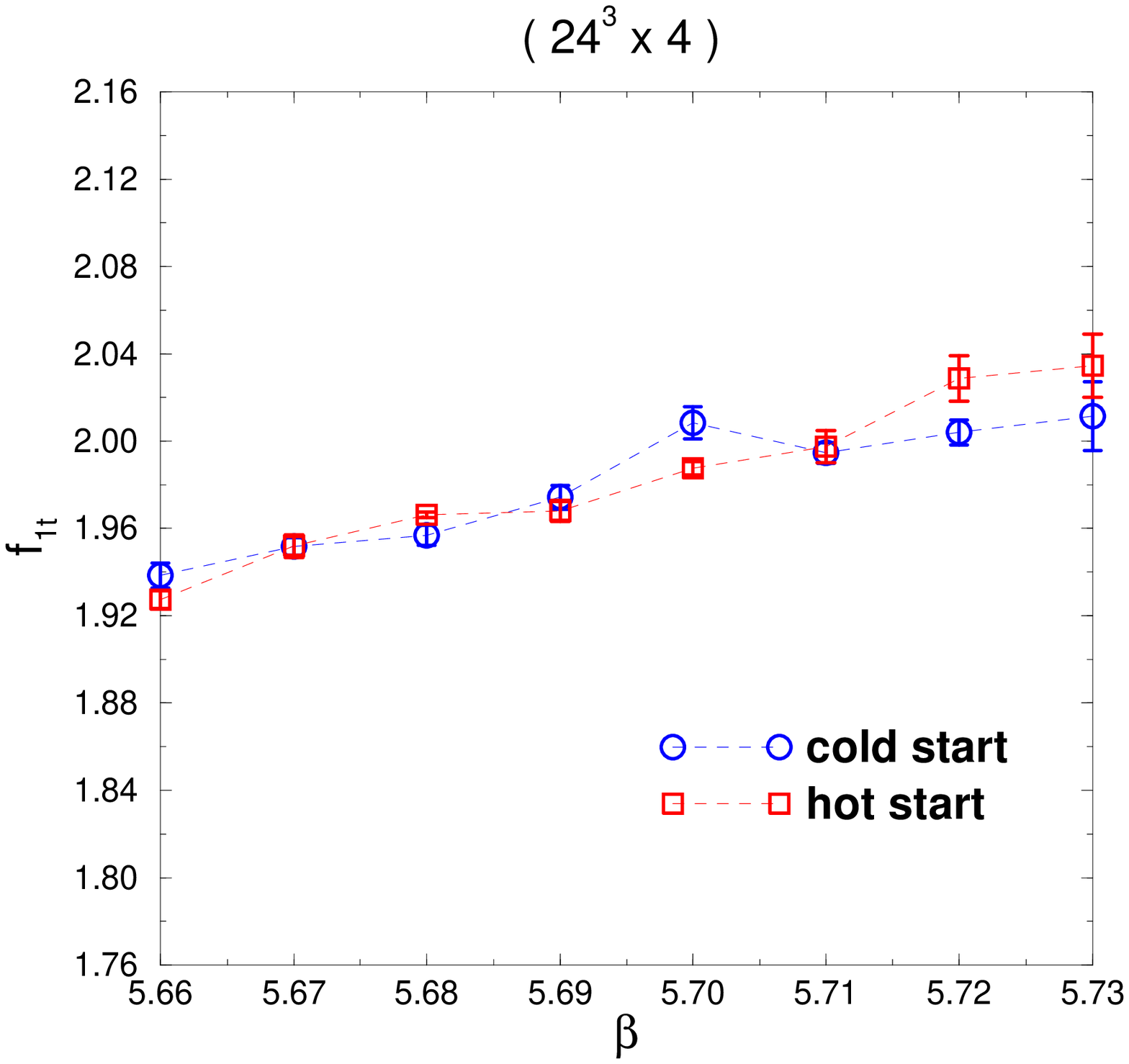}}
		\caption{
		$f_1$ for the time-like currents 
 near the transition temperature ($\beta_c = 5.69$).
		}
		\label{f1t}
	\end{figure}

The lattice sizes and $\beta$
considered are $24^3\times 4$ and 
 from $\beta=5.0$ to $\beta=6.3$ for the $T\ne 0$ case. 
Only the elementary monopoles are considered, since the time-extent is 
short. The monopole action also in this case is obtained beautifully.
\begin{enumerate}
\item
The action is calculated in both the confinement and in the 
deconfinement phases. Qualitatively the features are similar
 as in $T=0$ case.
$f_1$ is dominant in both phases.
In the deconfinement 
phase, however, there is a
 discrepancy between space-space and time-time coupling, 
whereas it is negligible in the confinement phase as seen in Fig.\ \ref{f1st}.
The critical $\beta$ is 5.69.
\item
Near the critical $\beta_c = 5.69$, we evaluated the monopole action
in detail. See Fig.\ \ref{f1s} and Fig.\ \ref{f1t} where 
the coupling of the self energy term connecting two space-like 
monopole currents 
and two time-like monopole currents are plotted respectively.
There is a clear hysteresis curve in Fig.\ \ref{f1s}. 
We could reproduce the first-order 
transition that is characteristic of the finite-temperature 
phase transition of pure $SU(3)$ QCD.
However such a clear hysteresis is not seen from the time-like currents.
This may be related to the fact that only the space-like currents are 
responsible for confinement as seen in \cite{ejiri95a,kita95}.
\end{enumerate}

\section{Conclusions}
We have derived 
an effective monopole action for various extended monopoles 
from vacuum configurations 
after abelian projection in the maximally abelian gauge in 
$T=0$ and $T\ne 0$ lattice $SU(3)$ QCD. 
We have restricted ourselves to the effective action for one type of 
monopole current after integrating out the other independent current.
The obtained results are very similar to 
those in $SU(2)$ case\cite{shiba4,shiba5}. 
The actions appear to be independent of the lattice volume. 
At zero temperature, 
monopole condensation is seen, 
for the first time in $SU(3)$, to occur from energy-entropy balance 
in the strong coupling region.
 Larger $\beta$ is included in the monopole condensed phase as more 
extended monopoles are considered. However, 
the scaling seen in the $SU(2)$ study is not yet observed. 
We have to study more block-spin transformations on larger lattices.
The renormalization flow diagram suggests the existence of an infrared 
fixed point where only a free theory exists.
A hysteresis behavior is seen around the critical 
temperature in the action of the $T\ne 0$ case.
Finally it is very important to derive the effective action for two 
independent monopole currents. It is our next target.

The simulations of this work were carried out on VPP500 at 
Institute of Physical and Chemical Research (RIKEN) and 
at National Laboratory for High Energy Physics at Tsukuba (KEK).
This work is financially supported by JSPS Grant-in Aid for 
Scientific  Research (B)(No.06452028).

\input epsf

%
%
	\newpage

	\newpage
	\newpage
	\newpage
	\newpage
	\newpage
	\newpage
	\newpage
	\newpage
	\newpage
%
%

\begin{references}
\bibitem{thooft1} G.'tHooft,{\it High Energy Physics}, ed.A.Zichichi
(Editorice Compositori, Bologna, 1975).
\bibitem{mandel} S. Mandelstam, Phys. Rep. {\bf 23C} (1976) 245.
\bibitem{poly} A.M. Polyakov, Phys. Lett. {\bf 59B}  (1975) 82.
\bibitem{bank} T.Banks $et\  al.$, Nucl. Phys. {\bf B129}  (1977) 493.
\bibitem{degrand} T.A. DeGrand and D. Toussaint, Phys. Rev. {\bf D22} 
 (1980) 2478.
\bibitem{thooft2} G. 'tHooft, Nucl. Phys. {\bf B190}  (1981) 455.
\bibitem{kron} A.S. Kronfeld $et\  al.$, Phys. Lett. 
{\bf 198B}  (1987) 516, \\
A.S. Kronfeld $et\  al.$, Nucl.Phys. {\bf B293}  (1987) 461.
\bibitem{yotsu} T. Suzuki and I. Yotsuyanagi, Phys. Rev. 
{\bf D42}  (1990) 4257;
Nucl. Phys. B(Proc. Suppl.) {\bf 20}  (1991) 236.
\bibitem{hio91} S. Hioki $et\  al.$, Phys. Lett. {\bf 272B}  (1991) 326.
\bibitem{suzu93} T. Suzuki, Nucl. Phys. 
B(Proc. Suppl.) {\bf 30} (1993) 176 
and references therein.  
\bibitem{wolo94} R.M. Woloshyn, Phys. Rev. {\bf D51} (1995), 6411.
\bibitem{miya95} O. Miyamura, Nucl. Phys. 
B(Proc. Suppl.) {\bf 42}  (1995) 538. 
\bibitem{wolo95} F.X. Lee et al., Nucl. Phys. 
B(Proc. Suppl.) {\bf 47}  (1996) 561. 
\bibitem{suzu95a} T.Suzuki et al., Nucl. Phys. 
B(Proc. Suppl.) {\bf 47}  (1996) 374. 
\bibitem{shiba94} H.Shiba and T.Suzuki, 
Phys. Lett. {\bf 333B} (1994) 461. 
\bibitem{stack94} J.D.Stack, R.J.Wensley and S.D.Neiman,
Phys.Rev. {\bf D50} (1994) 3399.
\bibitem{suzu95b} S. Ejiri et al., 
Nucl. Phys. B(Proc. Suppl.) {\bf 47}  (1996) 322. 
\bibitem{ejiri95a} S. Ejiri, Nucl. Phys. 
B(Proc. Suppl.) {\bf 47}  (1996) 539. 
\bibitem{peskin} M.E. Peshkin, Ann. Phys. {\bf 113}, (1978) 122.
\bibitem{frolich} J. Fr\"{o}lich and P.A. Marchetti, Euro. Phys. Lett.
{\bf 2},  (1986) 933.
\bibitem{smit} J. Smit and A.J. van der Sijs, 
Nucl. Phys. {\bf B355}, (1991) 603. 
\bibitem{shiba1} H.Shiba and T.Suzuki, 
Kanazawa University, Report No. Kanazawa 94-11, 1994. 
\bibitem{shiba2} H.Shiba and T.Suzuki, 
Kanazawa University, Report No. Kanazawa 93-09, 1993.
\bibitem{shiba3} H.Shiba and T.Suzuki, 
Nucl. Phys. B(Proc. Suppl.) {\bf 34}, (1994) 182.
\bibitem{shiba4} H.Shiba and T.Suzuki, 
Phys. Lett. {\bf B351} (1995) 519. 
\bibitem{shiba5} T.Suzuki et al.,
Nucl. Phys. B(Proc. Suppl.) {\bf 47}  (1996) 270. 
\bibitem{swendsen} R.H. Swendsen,Phys. Rev. Lett. 
{\bf 52}, (1984) 1165;
Phys. Rev. {\bf B30},  (1984) 3866, 3875.  
\bibitem{ivanenko} T.L. Ivanenko $et\  al.$, Phys. Lett. 
{\bf 252B},  (1990) 631.
\bibitem{suzu96} T.Suzuki et al., Talk at 'Lattice 96'.
To be published in Nucl. Phys. B(Proc. Suppl.).
\bibitem{schier} F. Brandstaeter et al., 
Phys. Lett. {\bf B272} (1991) 319.
\bibitem{shiba6} H.Shiba and T.Suzuki, 
Phys. Lett. {\bf B343} (1995) 315.
\bibitem{kita95} S. Kitahara, Y. Matsubara and T. Suzuki, 
Prog. Theor. Phys. {\bf 93} (1995), 1.

\end{references}
\end{document}